\documentstyle{article}
\def\abstract#1{\vskip 7mm 
        \begin{center}{\large Abstract}\par \vspace{12pt}
                \begin{minipage}[c]{12cm}
                        \small #1
                \end{minipage}
        \end{center}
}
\def\title#1{\begin{center}{\Large\bf #1}\end{center}}
\def\author#1{\vskip 5mm \begin{center}{#1}\end{center}}
\def\address#1{\begin{center}{\it #1}\end{center}}

\input epsf

\newcommand{\be}{\begin{equation}}
\newcommand{\ee}{\end{equation}}
\newcommand{\bea}{\begin{eqnarray}}
\newcommand{\eea}{\end{eqnarray}}

\newcommand{\bquote}{\begin{quote}}
\newcommand{\equote}{\end{quote}}
\newcommand{\lesson}[1]{\begin{quote}#1\end{quote}}

\newcommand{\eq}[1]{(\ref{eq:#1})}
\newcommand{\sect}[1]{Sec.~\ref{sec:#1}}
\newcommand{\fig}[1]{Fig.~\ref{fig:#1}}
\newcommand{\tabl}[1]{Table~\ref{table:#1}}

\def\IR{\relax{\rm I\kern-.18em R}}

\newcommand{\ap}{$\alpha'$}

\newcommand{\mr}[1]{\multicolumn{2}{r}{#1}}
\newcommand{\batu}{$\times$}
\newcommand{\tria}{$\triangle$}
\newcommand{\maru}{$\bigcirc$}
\newcommand{\arrows}{$\nearrow \quad \uparrow$}

\makeatletter
\@ifundefined{lesssim}{}{}
\@ifundefined{gtrsim}{}{}
\def\vereq#1#2{\lower3pt\vbox{\baselineskip1.5pt \lineskip1.5pt
\ialign{$\m@th#1\hfill##\hfil$\crcr#2\crcr\sim\crcr}}}
\makeatother

\begin{document}

\begin{flushright}
        KEK-TH-781\\		
        gr-qc/0108059\\	
\end{flushright}

\title{%
The Singularity Problem in String Theory
}
\author{Makoto Natsuume \footnote{E-mail: makoto.natsuume@kek.jp}}
\address{%
  Theory Division\\
  Institute of Particle and Nuclear Studies\\
  KEK, High Energy Accelerator Research Organization\\
  Tsukuba, Ibaraki, 305-0801 Japan
}

\abstract{
We review the current status of the singularity problem in string theory for non-experts. After the problem is discussed from the point of view of supergravity, we discuss classic examples and recent examples of singularity resolution in string theory. Based on lectures presented at ``Frontier of Cosmology and Gravitation" (April 25-27 2001, YITP), ``QFT2001" (July 16-19 2001, YITP), and the Fall Meeting 2001 of the Physical Society of Japan (Sep. 22-25 2001).
}

\section{Introduction}

In this note, we review the current status of the singularity problem in string theory, in particular emphasizing on the possible lessons one could draw from the known singularity resolutions in string theory.

The organization of the present note is as follows. First, in the next section, we discuss the singularity problem from the point of view of supergravity. Then in \sect{resolutions}, we see some ``classic" examples of singularity resolutions in string theory. However, we discuss in \sect{horowitz-myers} that there are at least two types of singularities --- one which is ``resolved" and the one which is ``excised." It is this latter type of singularities where we have a more understanding in the last few years. So, we quickly review these examples in \sect{recent}. In the course of discussion, we also discuss topology change and cosmic censorship conjecture in the context of string theory.

In the rest of this section, we briefly review supergravity -- the classical and the low energy approximations of string. However, it is not our purpose here to give a detailed review. For further discussion on supergravity, see the talk by Soda in this volume \cite{Soda:2001nz} and see, {\it e.g.}, Refs.~\cite{Big,Marolf:1998ci,Peet:2000hn}. A supergravity action derived from string theory typically has the following form:
\be
S = \frac{1}{16\pi G_d} \int d^{d}x \sqrt{-g} \left\{ e^{-2\phi} (R + 4(\nabla\phi)^2) + L_{matter} \right\},
\label{eq:string_metric}
\ee
where $G_d$ is the $d$-dimensional Newton's constant, $\phi$ is a scalar field called the dilaton, and $L_{matter}$ represents various matter fields such as gauge fields and fermions. If one makes a conformal transformation 
\be
g_{\mu\nu}^{E} = e^{-4\phi/(d-2)} g_{\mu\nu},
\label{eq:conformal}
\ee
then the action takes the standard Einstein-Hilbert form coupled to various matter fields:
\be
S_{E} = \frac{1}{16\pi G_d} \int d^{d}x \sqrt{-g^E} \left\{ R_{E} - \frac{4}{d-2}(\nabla\phi)^2 + L'_{matter} \right\}.
\label{eq:einstein_metric}
\ee
The metric $g_{\mu\nu}$ is called the string metric, and the metric $g_{\mu\nu}^E$ is called the Einstein metric. In this note, we will always write line elements $ds^2$ in the string metric.

Choosing a different conformal frame is just a field redefinition. Thus, physically they have to give equivalent answers. The definition of curvature  singularities, however, seems to depend on the conformal frame chosen since a metric changes under the field redefinition. In this sense, the notion of curvature singularities is not well-defined. But the dilaton is often singular even if a metric may become nonsingular in a given frame. Thus, one cannot really distinguish the singularity in the metric and the divergence of the dilaton. In either case, supergravity description is no longer valid near these ``singularities." (The Newton's constant corresponds to the constant part of the dilaton $e^{2\phi_0}$ from Eq.~\eq{string_metric}, so the quantum corrections become important once the dilaton becomes large.) We will call  ``singularities" whenever supergravity breaks down in such a broad sense.  

The action \eq{string_metric} is the leading order action, and it has various stringy corrections. For example, there are classical corrections from the massive string states (``\ap-corrections") such as
\be
l_s^2 \left\{ a_1 R_{\mu\nu\rho\sigma}R^{\mu\nu\rho\sigma} 
+ a_2 R_{\mu\nu}R^{\mu\nu} + a_3 R^2 \right\} + \cdots
\label{eq:alpha'}
\ee
with appropriate coefficients $a_i$, which depend on the theory one considers. Here, $l_s$ is the string scale, a typical size of string.

In general relativity, the event horizon of a stationary black hole must be topologically $S^2$. This is known as the topology theorem (Ref.~\cite{HE}, Proposition~9.3.2). One feature of higher dimensional theories is that the event horizon can take various kinds of topology \cite{Horowitz:1991cd}. These objects are generically called ``black $p$-branes." A black $p$-brane has the event horizon with topology $\IR^p \times S^{d-p-2}$. These branes arise because string theory has various $(p+1)$-form gauge potential $A_{p+1}$; a $p$-brane couples to $A_{p+1}$ and carries its charge. Again for more discussion on branes, consult other references.

For example, M-theory (the $d=11$ supergravity) and the Type IIA string has the branes in \tabl{mbrane}. 
\begin{table}[h]
\begin{center}
\begin{tabular}{llrrrr}
M-theory && W & M2 & M5 & KK \\
&&$\swarrow$\quad$\downarrow$&$\swarrow$\quad$\downarrow$&$\swarrow$\quad$\downarrow$&$\swarrow$\quad$\downarrow$ \\
Type IIA && D0 W & NS1 D2 & D4 NS5 & KK D6
\end{tabular}
\caption{M-theory branes and their relations to the Type IIA branes. NS$p$ represents a $p$-brane with a Neveu-Schwarz--Neveu-Schwarz (NS-NS) charge, and D$p$ represents a $p$-brane with a Ramond-Ramond (R-R) charge. W and KK represent the plane wave solution and the Kaluza-Klein monopole respectively.}
\label{table:mbrane}
\end{center}
\end{table}
Below we will mainly consider the BPS limit of black branes.

\section{Singularity Problem --- String as SUGRA}

\subsection{Singularity Theorem and Energy Conditions}\label{sec:energy_cond}

Most BPS branes in string theory are actually known to be singular (\tabl{mbrane2}). However, this fact itself is not very useful when one thinks about the singularity problem. First, one often knows only the solutions to the leading order action \eq{string_metric}, and the \ap-corrections may change the story. Moreover, when one talks of the singularity problem, it is not important that a particular solution has a singularity or not. What is important is that the occurrence of singularities is generically unavoidable or not. The singularity theorems are important in this respect in general relativity. Fortunately, many theorems in general relativity do not depend on the detailed form of the Einstein equation
\be
R_{\mu\nu} - \frac{1}{2} g_{\mu\nu} R = 8 \pi G T_{\mu\nu},
\label{eq:Einstein_eq}
\ee
{\it i.e.}, the detailed form of the energy-momentum tensor $T_{\mu\nu}$, but depend only on the generic conditions on $T_{\mu\nu}$. Thus, many results in general relativity can be extended rather easily to supergravity as well. Such conditions on $T_{\mu\nu}$ are generically named ``energy conditions." 

There are four energy conditions in common use:
\renewcommand{\arraystretch}{1.2}
\begin{center}
\begin{tabular}{lcl}
Condition & Definition & Applications \\
&&\\
Strong:& $ \left(T_{\mu\nu}-\frac{1}{d-2}g_{\mu\nu}T\right)t^{\mu}t^{\nu} \geq 0 $		& Hawking-Penrose singularity theorem \\
		&& (\cite{Hawking:1970sw}; Ref.~\cite{HE}, Sec.~8.2, Thm.~2) \\
Dominant:	& $ T_{\mu\nu} t^{\mu}\tilde{t}^{\nu} \geq 0 $	
			& positive mass theorem (ADM mass)  \cite{SC79a,SC79b} \\
			&& positive mass theorem (Bondi mass) 
			\cite{Horowitz:1982uw,Schoen:1982re} \\
			&& 0th law of black hole thermodynamics 
			\cite{Wald:1984rg} (p.333) \\
			&& topology theorem (no-hair theorem) \\
			&& (Ref.~\cite{HE}, Prop.~9.3.2; Ref.~\cite{nohair}, Thm.~6.17) \\
			&& cosmic censorship conjecture 
			\cite{Wald:1984rg} (p.303, p.305) \\
Weak:	& $ T_{\mu\nu} t^{\mu}t^{\nu} \geq 0 $ 
		& Penrose singularity theorem 
		(Ref.~\cite{HE}, Sec.~8.2, Thm.~1)\\
		&& 3rd law of black hole thermodynamics \cite{israel86} \\
		&& topology change (Tipler) \cite{Tipler:1977eb} (Thm.~5) \\
Null:	& $ T_{\mu\nu} n^{\mu}n^{\nu} \geq 0 $ 
		& 2nd law of black hole thermodynamics 
		\cite{Wald:1984rg} (Thm.~12.2.6) \\
		&& holographic $c$-theorem \cite{Freedman:1999gp}
\end{tabular}
\end{center}
\renewcommand{\arraystretch}{1}
where $t$ and $\tilde{t}$ are arbitrary future-directed timelike or null vectors, and $n$ is an arbitrary null vector. As other important applications, topological censorship \cite{Friedman:1993ty} and Penrose singularity theorem (Roman's improvement \cite{Roman:1986tp,Roman:1988vv}) use the averaged null energy condition, and Hawking-Penrose singularity theorem (Tipler's improvement) assumes the weak energy condition and the averaged strong energy condition \cite{tipler78}. One should note that the above applications do not only assume energy conditions, but also make various assumptions. From the above definitions, one can easily see that they are related to each other by 
\begin{center}
\begin{tabular}{ccc}
Strong			&				&		\\
				& $\searrow$	&		\\
				&				& Null	\\
				& $\nearrow$ 	&		\\
\begin{tabular}{ccc}
Dominant & $\rightarrow$ & Weak
\end{tabular}	&  				& 
\end{tabular}
\end{center}
($\rightarrow$ means that the tip of the arrow is the weaker condition).

There are many versions of singularity theorems, but the most powerful one is the Hawking-Penrose singularity theorem with the following assumptions (Ref.~\cite{Hawking:1970sw}; Ref.~\cite{HE}, Sec.~8.2, Theorem 2):
\begin{enumerate}
\item[(i).] The strong energy condition holds.
\item[(ii).] There exists a compact surface called a trapped surface, at which all null geodesics orthogonal to the surface converge. 
\item[(iii).] No closed timelike curve exists.
\item[(iv).] The timelike and null generic conditions are satisfied.
\end{enumerate}
Thus, as long as these conditions are satisfied, singularities are expected to occur generically for any metric theories of gravity. Below we will consider the strong energy condition.

The strong energy condition holds for massless scalar fields or for gauge fields, but even physically reasonable cases violate the condition. For example, for a massive scalar field $\varphi$, 
\be
\left(T_{\mu\nu}-\frac{1}{d-2}g_{\mu\nu}T \right) 
t^{\mu}t^{\nu} 
= (t \cdot \nabla\varphi)^2 + \frac{1}{d-2} m^2 \varphi^2 t^2 \not\geq 0.
\ee
The first term is positive definite, but the second term can be negative since $t^2<0$. Similarly, a positive scalar potential violates the condition too. Since a scalar potential contributes to a cosmological constant, a positive cosmological constant violates the condition as well; thus, the condition is violated in inflationary universes.

In string theory, the dilaton violates the condition in the string metric. Also, string theory has corrections such as \eq{alpha'}. Because the curvature becomes large near a singularity, such corrections give large corrections to a solution. Formally regarding such corrections as $T_{\mu\nu}$, one can see that such corrections violate the strong energy condition as well. Thus, the singularity theorem does not hold in string theory. There could be black holes without singularities. (Incidentally, there have been known some special black holes without singularities by violating the assumptions other than (i). The M5-brane violates the condition~(ii) (the trapped surface is not compact since it is a brane) \cite{Gibbons:1995vm}, and $J \neq 0$ BTZ black holes violate the condition~(iii).)

Since the curvature becomes large near singularities, singularities might be resolved by \ap-corrections. However, in practice, it is hard to check this explicitly since the correction~\eq{alpha'} is only the leading order correction. Once the leading correction becomes important, all the corrections had better be included, which are practically impossible.

\subsection{Classically Singular Spacetimes in String}\label{sec:exact}

However, there have been known many exact solutions with all \ap-corrections, and some of them are singular. Among the BPS brane solutions in string theory, the NS1-brane and the plane wave are known to be exact solutions with singularities. The NS1-brane is given by \cite{Dabholkar:1990yf}
\bea
ds^2 &=& f^{-1} (-dt^2+dx^2) + dr^2 + r^2 d\Omega_7^2, \nonumber \\
e^{-2\phi} &=& f, \\
B_{tx} &=& f^{-1}-1, \nonumber
\eea
where the brane lies along $x$ direction, $f=1+(r_0/r)^6$, and $r=0$ is the singularity. This solution was initially obtained without including \ap-corrections, but is known to be exact even with all classical corrections \cite{Horowitz:1994ei}.

The plane wave solution is an exact classical solution of any metric theory of gravity (as long as the theory reduces to general relativity at low energy) \cite{Gueven:1987ad,Horowitz:1990bv}: 
%
\be
ds^2 = - du dv + dx^i dx_i + F(u,x^i) du^2, \quad \partial_i^2 F=0,
%
\ee
where $i$ runs from 1 to $d-2$. The reason is because the solution has a  covariantly constant null vector $l$, {\it i.e.},
\be
\nabla_{\mu} l^{\nu}=0, \quad l^2=0,
\label{eq:killing}
\ee
and the Riemann tensor is proportional to two powers of $l$. Since any contraction of $l$ vanishes, all second-rank tensors constructed from curvature tensors and their derivatives must vanish. So, all \ap-corrections to the equation of motion vanish identically. Similarly, all curvature invariants vanish as well. However, it still has a singularity if $F$ diverges. This is because the tidal forces diverge as one approaches the singularity. In general relativity, one can classify singularities as follows \cite{tipler80,HE,Wald:1984rg}: 
\begin{enumerate}
\item s.p. (scalar polynomial) curvature singularities; 
\item p.p. (parallelly propagated) curvature singularities; 
\item others (the conical singularity, Taub-NUT space, \ldots).
\end{enumerate}
The singularity of the plane wave corresponds to 2. 

There is no known generic singularity theorem which applies to string theory. However, since there are many classically exact solutions with singularities, 
\lesson{Lesson 1: String (as a classical field theory) does not prohibit singularities.}
When one talks of singularities in string theory, one should really look at classically exact solutions. However, below we will mainly focus on the leading order solutions for simplicity.

\subsection{Embeddings}\label{sec:embedding}

As a classical field theory, string does not prohibit singularities. Thus, one needs some stringy mechanism to resolve singularities. However, some singularities can be resolved simply by embedding into higher dimensions \cite{Gibbons:1995vm}. 

For example, the $d=10$ Type IIA string is embedded into the $d=11$ supergravity as 
\be
ds_{11}^2 = e^{-2\phi/3} ds^2 + e^{4\phi/3}(dx^{11}+A_{\mu}dx^{\mu})^2,
\label{eq:embedding}
\ee
where $ds_{11}^2$ and $ds^2$ are the line elements in 11 and 10-dimensions respectively (The latter is the line element in the string metric). 

As seen from \tabl{mbrane2}, most BPS branes in the Type IIA string are singular (in the leading order solutions). The NS5-brane is nonsingular, but the string coupling diverges as one approaches the horizon. Thus, supergravity again breaks down there. In this sense, the NS5-brane is singular as well.

Most Type IIA branes are singular, but some of them can be resolved by embedding into M-theory. Since M-theory is regarded as  more fundamental than string theory in recent years, brane singularities are not very serious in this respect. However, obviously this method is not enough. Even M-theory has singular solutions, and this method  cannot be applied to the Type IIB string, where there is no embedding in higher dimensions. But one should note that some singularities can be resolved without an exotic mechanism.

Even though NS1 and W are exact solutions, they are vacuum solutions and lack source terms; they describe only the ``Coulomb part" of the field. Including the source term in the field equation may lead to \ap-corrections and may smooth out these singularities \cite{Tseytlin:1995uq}. 

\begin{table}
\begin{center}
\begin{tabular}{llcccccccr}
M-theory	&& \mr{W}		& \mr{M2}		& \mr{M5}		& \mr{KK}		\\
\quad Singularity 
			&& \mr{\tria}	& \mr{\batu}	& \mr{\maru}	& \mr{\maru}	\\
			&& \mr{\arrows}	& \mr{\arrows}	& \mr{\arrows}	& \mr{\arrows}	\\
Type IIA	&& D0 & W		& NS1 & D2		& D4 & NS5		& KK & D6	\\
\quad Singularity (string metric) 
			&& \tria&\tria	&\batu &\tria	&\batu &\maru	& \maru&\batu\\
\quad Singularity (Einstein metric) 
			&& \batu&\tria	&\batu &\batu	&\batu &\batu	& \maru&\batu\\
\quad Effective coupling
			&& S    & $-$	& W    & S	& W    & S	& $-$  & W 
\end{tabular}
\caption{Singularities for M-theory and Type IIA branes (BPS limit). The definition of a curvature singularity depends on the conformal frame chosen. Thus, the presence of singularities are shown both in the string metric and in the Einstein metric. We also show how the effective coupling $e^{\phi}$ varies as one approaches the branes. The table below shows what various symbols stand for.}
\label{table:mbrane2}
\vspace{5mm}
\begin{tabular}{|l|c|l|} 
\hline \hline
Singularity & \maru & regular \\
	& \batu & s.p. curvature singularity \\
	& \tria & p.p. curvature singularity \\ \hline
String coupling & W & weak coupling \\
	& S & strong coupling \\
	& $-$ & constant \\ 
\hline \hline 
\end{tabular}
\end{center}
\end{table}

\section{Singularity Problem --- Stringy Resolutions}\label{sec:resolutions}

\subsection{Dualities}\label{sec:tdual}

As we saw, classical spacetime itself could be singular even in string theory. However, this does not necessarily imply that physics is singular there. The natural probe in string theory is string and branes. If physics of string and branes itself is not singular, a singularity in a metric does not matter. 

\begin{figure}
\begin{center}
\hspace{.2in}\epsfbox{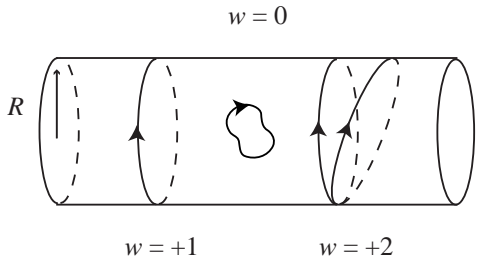}
\caption{Closed strings of winding number $w=+1, 0, +2$.}
\label{fig:cylinder}
\end{center}
\end{figure}

Strings behave very differently from point particles. As a simplest example, let us consider strings on a cylinder with radius $R$ (\fig{cylinder}). The effect of the compactification appears in two ways. First, the center-of-mass momentum is quantized:
\be
k=\frac{n}{R}.
\ee
This is just a Kaluza-Klein momentum and is the same as in field theory. The second effect is stringy; a closed string may wrap around the cylinder. Since the string has a tension, its potential energy is proportional to the total length of the string:
\be
E=\frac{wR}{l_s^2}.
\label{eq:winding}
\ee

Now, the spectrum is invariant under
\be
R \leftrightarrow\frac{l_s^2}{R}, \qquad n \leftrightarrow w.
\ee
This symmetry is called a T-duality. (See Ref.~\cite{Giveon:1994fu} for a review.) Since the spectrum is equivalent, we cannot distinguish the cylinder with radius $R$ and the cylinder with radius $l_s^2/R$. Thus, the physically inequivalent theories are limited to $R \geq l_s$. This implies that string theory has a ``minimal length" $l_s$. (If one considers branes, there could be a shorter distance scale though.) 

The T-duality for the cylinder $R \leftrightarrow l_s^2/R$ can be extended to more general spacetimes. Let translations in $x$ be a Killing symmetry. In the simplest case, if the original solution is $(g_{\mu\nu}, \phi)$,
the transformed solution $(\tilde{g}_{\mu\nu}, \tilde{\phi})$ is given by
\be
\tilde{g}_{xx}=1/g_{xx}, \qquad \tilde{\phi}=\phi-\frac{1}{2}\ln g_{xx}
\label{eq:tdual}
\ee
(See, {\it e.g.}, Ref.~\cite{Rocek:1992ps} for the full T-duality transformation). As an example, consider four-dimensional Minkowski spacetime in cylindrical coordinates:
\be
ds^2 = - dt^2 + dx^2 + dr^2 + r^2 d\theta^2.
\ee
The duality transformation to $\theta$ changes $g_{\theta\theta}$ to $r^{-2}$ and creates a curvature singularity at $r=0$. But the theory on this spacetime should be equivalent to the original theory on flat spacetime. Thus, the would-be singularity is not a true singularity, but merely an artifact of classical field theory. For instance, the T-duality rule~\eq{tdual} can get \ap-corrections. This is a simple example, but there are other nontrivial examples, where a singular spacetime transforms to a nonsingular spacetime via T-duality ({\it e.g.}, see Ref.~\cite{Horowitz:1993jc}).

Strictly speaking, T-duality requires that $x$ is compact and the signature of the metric is Euclidean \cite{Rocek:1992ps}. Otherwise, the symmetry is currently justified only as a symmetry of supergravity and maps solutions to solutions. (To begin with, many issues have not been solved for string theory on curved spacetime, {\it e.g.}, see Refs~\cite{Asano:2000fp,Natsuume:1998ij}.)

There are many attempts to resolve singularities using other dualities. Ref.~\cite{Itzhaki:1998dd} uses gauge/gravity dualities for D-branes in order to resolve singularities of these branes. We will see other examples in \sect{recent}, which makes use of this duality.

One sometimes seems to have the impression that duality arguments ``sweep the problem under the rug." Even though the problem is solved in the dual picture, it is not clear how the problem is solved in the original picture. In this sense, duality arguments are more like the existence proof of the solution. It guarantees that there is a solution, but does not tell how it is solved. However, duality arguments are natural logically. Recall that supergravity is only an approximation to string theory. String has various approximation schemes ({\it e.g.}, supersymmetric Yang-Mills in gauge/gravity duality) depending on the situations. The existence of singularities simply tells us that supergravity is no longer a valid description due to corrections such as \eq{alpha'}. In this case, it is natural to use more appropriate scheme; one should not stick to an approximation which is not valid.

\subsection{Orbifolds}\label{sec:orbifold}

\begin{figure}
\begin{center}
\hspace{.25in}\epsfbox{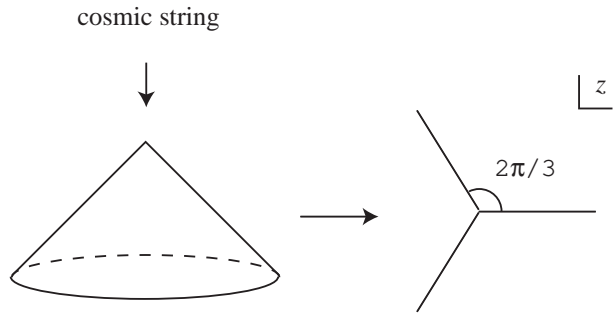}
\caption{Infinitely thin cosmic string. For illustration, only the transverse directions to the cosmic string are shown.}
\label{fig:cosmic_string}
\end{center}
\end{figure}

The other well-known example of singularity resolution is orbifolds \cite{orbifold}. An example of orbifolds is an infinitely thin cosmic string \cite{Vilenkin:1985ib}. Let us consider a cosmic string with a deficit angle $4\pi/3$ (\fig{cosmic_string}). The transverse plane to the cosmic string is described by a ``cone," but one can instead consider a flat space and identify three regions. Denoting the transverse plane by a complex variable $z$, we identify $z \rightarrow \exp(2\pi i/3)z$. As the result of the identification, this theory has a $Z_3$ symmetry, so one can write
\be
\IR^2/Z_3.
\ee
The tip of the cone $z=0$ is a singularity. The curvature does not diverge there since we simply identify the plane. But the point $z=0$ is not a smooth manifold, and the spacetime is geodesically incomplete there. In the classification of singularities in \sect{exact}, the conical singularity is one type of 3, called quasiregular singularity.

A field theory on such a background is generically problematic. The Hilbert space on the orbifold is obtained from the flat space Hilbert space by truncating $Z_3$-noninvariant states. However, the $Z_3$-noninvariant states truncated may be indispensable for unitarity. If they are not decoupled from the rest of the spectrum, we would spoil unitarity by the truncation. 

\begin{figure}
\begin{center}
\hspace{0.5in}\epsfbox{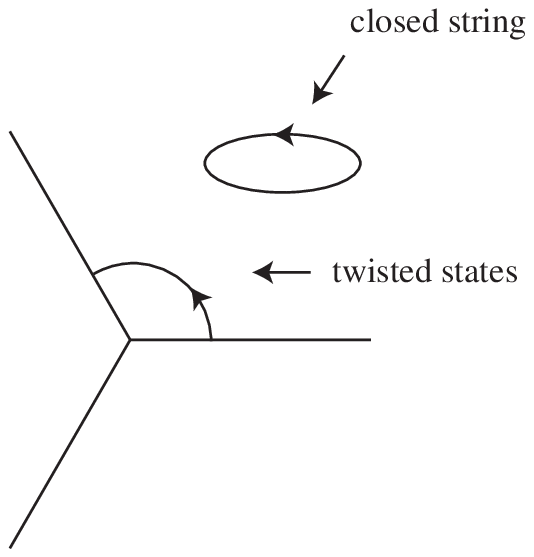}
\caption{}
\label{fig:twisted}
\end{center}
\end{figure}

On the other hand, string theory on orbifolds are unitary. The Hilbert space of string is not only the $Z_3$-invariant subspace, but there are also new sectors. Because of the $Z_3$ symmetry, it is enough for closed strings to close up to $Z_3$ action (\fig{twisted}). These strings are called the twisted sectors, and they correspond to the winding strings in \sect{tdual}. With the addition of these new sectors, unitarity is recovered. String theory on orbifolds is perfectly consistent, so the singularity does not matter. To be precise, ordinary orbifolds are compact ones. A noncompact orbifold such as $\IR^2/Z_3$ has interesting new physics in terms of singularity resolution \cite{Adams:2001sv}.

One common theme both for T-duality and for orbifolds is that the extended nature of strings is essential for resolving singularities. Thus, one lesson here is
\lesson{Lesson 2: Pay attention to the extended objects.}
The theme will appear below over and over again. Also, most of discussion so far is essentially classical. So,
\lesson{Lesson 3: Some singularities are resolved in string theory even ``classically."}
This is because string theory has new physics even classically ({\it e.g.}, winding modes).

\subsection{Topology Change}\label{sec:topology_change}

\begin{figure}
\begin{center}
\hspace{0.25in}\epsfbox{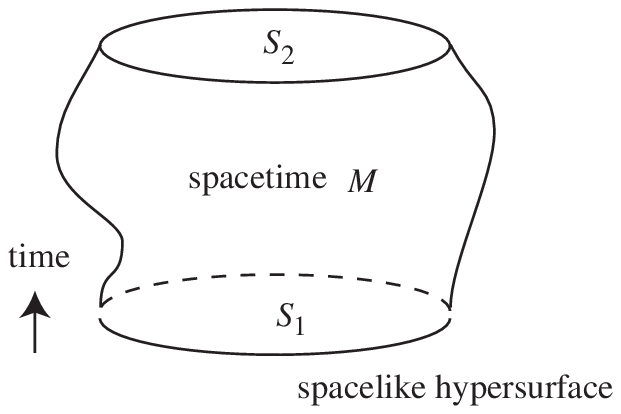}
\caption{}
\label{fig:topology_change}
\end{center}
\end{figure}

Now, let us move on the issue of topology change. We include the issue here since topology change is always associated with singularities in general relativity; smooth topology change cannot occur in general relativity. Suppose that a spacetime $M$ has two spacelike hypersurfaces $S_1$ and $S_2$ which have different topology (\fig{topology_change}). Then, the spacetime $M$ must have either closed timelike curves or singularities \cite{geroch67}. The argument is very generic, and does not depend on any equations of motion. However, if one imposes the weak energy condition and the generic condition, a stronger statement holds; the spacetime must be actually singular \cite{Tipler:1977eb}. Now, does smooth topology change occur in string theory?

Compactify spacetime to four-dimensional Minkowski spacetime and a six-dimensional compact Calabi-Yau space. We will consider topology change in the compact space. 
To low energy observers, topology change looks as a phase transition because additional particles become massless, and the symmetry breaking pattern changes at the point of the topology change. In string theory, the spectrum of massless particles is determined by the topology of the compact manifold. For example, the number of generations is given by $|\chi (K)|/2$, where $\chi (K)$ is the Euler number of the compact space $K$. Two kinds of topology change have been known: flops \cite{AGM,Witten:1993yc} and conifold transitions \cite{Strominger:1995cz,Greene:1995hu}. (See Ref.~\cite{Greene:1996cy} for a review.)

The problem of conifold transitions was particularly keen in the Type IIB string. When the topology change occurs, $S^3$ (part of the compact space) shrinks to  zero size at the singular point. But this theory has D3-branes. The D3-brane which wraps around $S^3$ becomes massless at the point (since the mass of the brane is the product of the brane tension and its volume). Low energy physics looks singular since this new massless state was not taken into account. 
Also, the massless state can condense and gives mass to the other massless fields. Geometrically, the whole process looks as the topology change from a compact space with $S^3$ to the one with $S^2$. The discussion of flops uses mirror symmetry, which is a generalization of T-duality. Even though the transition looks singular in the original picture, it is completely a smooth change in the mirror description. We see earlier that extended objects are important to resolve singularities, but this is because

\lesson{Lesson 4: A singularity signals the appearance of new massless degrees of freedom.}

\section{Resolutions and Excisions}\label{sec:prohibited}

\subsection{Singularities Which Are Not Resolved}\label{sec:horowitz-myers}

So far we have seen various mechanisms to resolve singularities. 
These singularities are basically well-defined, and one can interpret them as physically reasonable sources. Such a singularity may be called as a ``geometric singularity." While they have a wide range of applications in string theory, geometric singularities are only a part of singularities that could arise, and not all singularities have a reasonable interpretation quantum mechanically.
Namely, not all singularities can be resolved \cite{Horowitz:1995ta}. 

As an example, consider the $M<0$ Schwarzschild solution. Asymptotically the metric is a good approximate solution even in string theory. It is not clear what happens near the singularity $r=0$. It could be possible that the singularity is resolved by \ap-corrections and that the full solution is nonsingular. But then the theory would have a regular solution with negative energy, so Minkowski spacetime would not be stable. Even if the theory admits a stable ground state with $E<0$, we can argue similarly by starting with the Schwarzschild solution with $M<E$. Thus, if all singularities are resolved, states with arbitrarily negative energy are allowed, and there is no guarantee that a ground state exists.

Therefore, some singularities should not be resolved, but must be prohibited somehow. References~\cite{Gubser:2000nd,Maldacena:2001mw} propose what kinds of singularities must be prohibited. Now, the $M<0$ Schwarzschild considered here is a naked singularity, and the statement that the $M<0$ Schwarzschild must be prohibited may sound familiar to relativists --- there is a well-known cosmic censorship conjecture which protects the occurrence of such a naked singularity. In fact, for the $M<0$ Schwarzschild, its gravitational force is repulsive; massive particles are repelled. It is hard to imagine how such a singularity can actually form from gravitational collapse. Thus, one candidate to prohibit unphysical singularities is the cosmic censorship, and we will see its status in string theory below.

In \sect{recent}, we will encounter such a singularity in the context of string theory, but with a twist. Many such solutions are BPS solutions composed of elementary branes. As a result, even if one cannot interpret the singularities as physically reasonable sources, the brane configuration themselves must exist. It seems that string theory solves the issue by changing the geometries drastically (by some stringy effect) so that the resulting geometries are well-behaved.

\begin{figure}
\begin{center}
\hspace{0.5in}\epsfbox{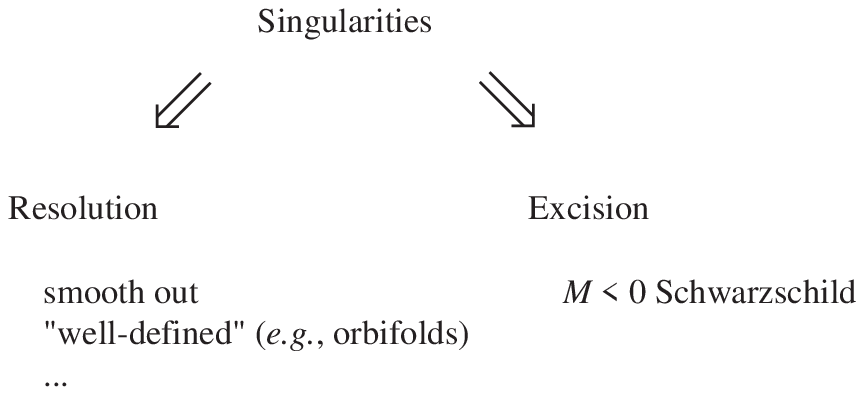}
\caption{}
\label{fig:2class}
\end{center}
\end{figure}

We have seen that there are two classes of singularities (\fig{2class}). 
One is ``resolved" singularities; they are, {\it e.g.}, ``well-defined" or smoothed out by \ap-corrections. The other is ``excised" singularities; string theory had better prohibit them as unphysical. We will loosely call both mechanisms as ``resolutions." 

\subsection{Cosmic Censorship}\label{sec:cosmic_censorship}

There are two versions of the cosmic censorship conjecture. Roughly speaking, they state as follows:
\bquote
Weak Cosmic Censorship Conjecture \cite{Penrose:1969pc}: Starting from generic initial conditions, all singularities from the gravitational collapse of physically reasonable matter are hidden within black holes in asymptotically flat spacetimes.
\equote
\bquote
Strong Cosmic Censorship Conjecture \cite{Penrose:1980ge}: For all physically reasonable spacetimes, no singularity is ever visible to any observer apart from a possible initial singularity (such as the big bang singularity), {\it i.e.}, timelike singularities never occur.
\equote
The strong censorship in particular implies that singularities are not only hidden to asymptotic observers, but also to the observers who fall into black holes. 

Below we will consider only the weak censorship. The conjecture is essential to derive various properties of black holes ({\it e.g.}, the second law of black hole thermodynamics, the positive mass theorem, and the topology theorem). However, there was a possible counterexample until recently that higher dimensional theories like string theory would violate the conjecture. 
What is special to these theories is the existence of branes, and this fact leads to the would-be counterexample. 
(See Ref.~\cite{Wald:1997wa} for the current status of the cosmic censorship conjecture in general relativity.)

To begin with, black branes are unstable under perturbations, which is known as the Gregory-Laflamme instability \cite{GL}. Since this phenomenon itself has wide applications, we describe it briefly. The simplest example occurs for neutral objects in five dimensions (with $S^1$ compactified. The $S^1$ is just an infrared cutoff.) The instability occurs since a neutral solution in five dimensions is not unique and there exist at least two. One is a $S^1$-wrapped black string:
\be
ds^2 = -\left(1-\frac{r_0}{r}\right) dt^2 + \frac{dr^2}{1-\frac{r_0}{r}} + r^2 d\Omega_2^2 + dx^2.
\ee
Here, $x$ is $S^1$ direction with circumference $L$. This is simply the product of a $d=4$ Schwarzschild and $S^1$. The $d=4$ Newton's constant $G_4$ is related to the $d=5$ Newton's constant $G_5$ by $G_4=G_5/L$. This is because the four-dimensional reduced action then takes the standard Einstein-Hilbert action
\be
\frac{1}{16\pi G_5} \int d^{5}x \sqrt{-g_5} R_5 
= \frac{1}{16\pi G_4} \int d^{4}x \sqrt{-g_4} \{ R_4 + \cdots\},
\ee
where the dots denote matter fields that arises from the compactification. Thus, the Bekenstein-Hawking entropy of the black string is given by
\be
S_{BH}^{(string)} \sim \frac{A_4}{G_4} \sim \frac{r_0^2}{G_4} \sim G_4 M^2 \sim \frac{G_5 M^2}{L},
\ee
where $M$ is the mass of the black string. We have derived the entropy from the point of view of four-dimensional observers, who do not see the fifth dimension. But the entropy can be calculated from the point of view of five-dimensional observers with the same result. Using $G_4=G_5/L$, and that the $d=4$ area $A_4$ is related to the $d=5$ area $A_5$ by $A_5=A_4 L$, one gets $A_5/G_5 = A_4 L/ G_4 L = A_4/G_4$.

Another neutral configuration is the black hole solution in $\IR^{1,3} \times S^1$. Since we will concern with the situation where $L$ is large, the solution is well-approximated by the $d=5$ Schwarzschild solution:
\be
ds^2 \sim -\left\{1-(\frac{\tilde{r}_0}{r})^2\right\} dt^2 + \frac{dr^2}{1-(\frac{\tilde{r}_0}{r})^2} + r^2 d\Omega_3^2.
\ee
The entropy of the black hole is given by
\be
S_{BH}^{(BH)} \sim \frac{A_5}{G_5} \sim \frac{\tilde{r}_0^3}{G_5} \sim \sqrt{G_5} M^{3/2}.
\ee

\begin{figure}
\begin{center}
\hspace{.25in}\epsfbox{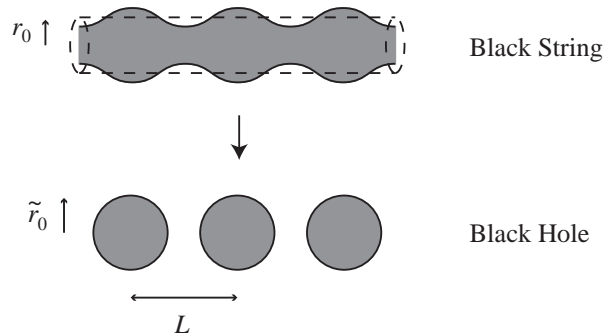}
\caption{A black string is unstable under perturbations (The Gregory-Laflamme instability). One possible end-product of the instability is a black hole.}
\label{fig:cosmic_censorship}
\end{center}
\end{figure}

For large enough $L$, $S_{BH}^{(BH)}>S_{BH}^{(string)}$. So, the black string is expected to be unstable and decay. One possible outcome would be the black hole. However, an event horizon does not bifurcate under the cosmic censorship (Ref.~\cite{Wald:1984rg}, Theorem~12.2.1). Thus, a naked singularity might arise if the decay really happens (\fig{cosmic_censorship}).

Now, we make a couple of remarks:
\begin{itemize}

\item The black string discussed here is not a BPS solution like D-branes. BPS solutions are expected to be stable from supersymmetry, and their analysis shows that this is in fact the case. Also, in principle this instability is not the instability associated with the Hawking radiation, but there is an interesting relation between them \cite{GM}.

\item This instability does not occur in general relativity. The black string which approaches Minkowski spacetime asymptotically does not exist in four dimensions under the strong/weak/dominant energy conditions \cite{Horowitz:1991cd}. This is also a consequence of the topology theorem.


\end{itemize}

There are many possible loopholes in the argument, so one cannot immediately say that the censorship is violated. In fact, Ref.~\cite{Horowitz:2001cz} has shown that the event horizon does not bifurcate under the instability. So, the Gregory-Laflamme instability does not seem to end up with black holes. Actually, there is no reason to believe that the above two are the only possible neutral configurations . They conjecture the existence of the {\it stable} black string which breaks the translational invariance (along the string). They claim that this black string is the end-product of the instability. The Gregory-Laflamme instability was the only known counterexample that would violate the censorship in string theory. Since this possibility now seems gone, there is no indication that the (weak) cosmic censorship is violated even in string theory.


\subsection{Recent Examples}\label{sec:recent}

Recently, many new examples are found to resolve singularities. References~\cite{Johnson:2000qt}-\cite{Adams:2001sv} are only a partial list. They typically resolve naked singularities.

Many of them are systems with D3-branes, and they have fewer supersymmetry than the maximal one. 
%
%
This is because they originated from the study of gauge/gravity duals. 
The original Maldacena duality was the following duality (See Ref.~\cite{Aharony:2000ti} for a review about the duality.):
\begin{center}
Type IIB string on $AdS_5 \times S^5$ $\leftrightarrow$ $D=4$, ${\cal N}=4$ supersymmetric Yang-Mills theory
\end{center}
It is of course very important to study other duals with fewer supersymmetry in order to study more realistic gauge theory and its physics ({\it e.g.}, confinement). This is obviously a topic many people have thought, and thus we have many examples. In this case, the background is often $AdS_5 \times X$ asymptotically, where $X$ is a more general transverse space than the original $S^5$. So, naive deformations of $AdS_5$ tend to produce singularities, but stringy physics cures the problem. 

The singularity in the original geometry is often naked. Moreover, they are ``unphysical." A typical unphysical behavior is that gravitational force becomes repulsive near the singularity. This type of singularities is known as ``repulson." (The extreme Reissner-Nordstr\"{o}m solution has the same behavior inside the degenerate horizon.) They are unphysical just like the $M<0$ Schwarzschild, but the final geometries usually do not have the problem ({\it e.g.}, all enhan\c{c}on examples known to date, the Klebanov-Strassler solution, and some of ``transgression" examples.) In the terminology of \sect{horowitz-myers}, they ``excise" singularities, not ``resolve" singularities. 

Now, we briefly review a couple of examples:
\begin{enumerate}
\item[(i).] Enhan\c{c}on mechanism \cite{Johnson:2000qt}: 
The prototypical example is D$p$-branes wrapped around K3 ($p = 4, \ldots, 7$), and the supergravity solutions have repulson singularities.
%
%
However, the constituents, the wrapped D-branes, become massless before reaching the singularity. (This is similar to the conifold transition in \sect{topology_change}, but the new massless degrees of freedom appear {\it before} the singularity, not at the singularity.) Thus, the naive supergravity analysis breaks down, and one must take into account stringy physics. Their conclusion is that constituent branes stay at a finite radius and form a shell (called enhan\c{con}). 
%
%
This mechanism is closely related to T-duality in \sect{tdual}. The wrapped D4-branes are nothing but winding states in a dual picture \cite{Natsuume:2001qt} under the $d=6$ heterotic/IIA S-duality \cite{Hull:1995ys,Witten:1995ex}. The gauge theory side is pure ${\cal N}=2$ supersymmetric Yang-Mills theory. The mechanism has the gauge theory interpretation that the metric on moduli space is one-loop exact but receives corrections nonperturbatively. Thus, the phenomenon seems characteristic to an ${\cal N}=2$ system. For more details, see the talk by Yamaguchi in this volume \cite{Yamaguchi:2001yd}. Also, see Refs.~\cite{Buchel:2001cn}-\cite{Merlatti:2001gd} for this mechanism applied to other systems. 
\item[(ii).] Polchinski - Strassler solution \cite{Polchinski:2000uf}: On the gauge theory side, one may add mass perturbations to the ${\cal N}=4$ matter in order to break supersymmetry (so-called ${\cal N}=1^*$ theory). Then, the gravity side looks singular \cite{Girardello:2000bd}, but actually it is not due to Myers' effect (dielectric effect) \cite{Myers:1999ps}. A collection of D-branes develops a multipole moment under a background field that normally couple to a higher-dimensional brane. This effect arises because D-brane coordinates become noncommutative for coincident D-branes. For example, D0-branes in an RR 4-form flux blow up into a spherical D2-brane, and develops a dipole moment under the potential. Similarly, in the Polchinski - Strassler solution, D3-branes in a RR 7-form flux polarize the D3-branes into a D5-brane. The RR 7-form flux corresponds to the mass perturbations. Both the enhan\c{c}on mechanism and Myers' effect could be considered as some sort of ``brane expansion phenomena," and such a finite distribution of branes resolves singularities. There is another example called the ``giant gravitons" \cite{McGreevy:2000cw}. Reference~\cite{Myers:2001aq} uses the giant gravitons to resolve a certain naked singularity.
They are all related via appropriate dualities (For a review, see Ref.~\cite{Myers:2001ks}).
\item[(iii).] Klebanov - Strassler solution \cite{Klebanov:2000hb}: The system consists of D3-branes and fractional D3-branes on the conifold. The fractional D3-branes are D5-branes wrapped over vanishing $S^2$. The solution is known to be singular \cite{Klebanov:2000nc}, but the solution does not have the right symmetry expected from gauge theory. On the gauge theory side, one has a ${\cal N}=1$ theory. The theory has an anomalous $U(1)_R$ symmetry, which is broken down to $Z_2$ by nonperturbative effect. On the gravity side, replacing the conifold by the ``deformed" conifold does the job, and the resulting geometry is regular. In the final analysis these constituent branes dissolve into flux. There is another way to modify the conifold (``resolution"), but this manipulation is not enough to repair the geometry and may need the enhan\c{c}on mechanism as well \cite{PandoZayas:2000sq}. Also, this solution has been used for a stringy realization of the Randall-Sundrum model (RS1) \cite{Giddings:2001yu}. For other related issues, see Refs.~\cite{Maldacena:2001mw}-\cite{Atiyah:2001qf}.
\end{enumerate}

Although many examples are known, the underlying stringy physics is not often very clear. Typically, there are two problems:
\begin{itemize}
\item Gauge theory side picture only: One often has the understanding in the gauge theory side only. This problem is common to duality arguments. However, this itself is fine if the gravity side is strongly-coupled and the notion of geometry does not make sense any more ({\it e.g.}, the case of Ref.~\cite{Itzhaki:1998dd}). But since the original geometry has a large unphysical region, 
%
%
the theory repairs the region where geometry still makes sense. 
\item Just replace the singular geometry with a regular geometry: Some approaches declare that the original solution is simply wrong, and they just replace it with a regular geometry. Note that there is no no-hair theorem here since we are talking of naked singularities. The solution may not be unique for a given set of conserved charges.
\end{itemize}
Many examples use both approaches simultaneously to make scenarios convincing. Also, in many examples, the backgrounds are often very complicated, and few exact solutions are known (even the leading order solution in \ap).

\section{Conclusions}

We saw that singularities are often resolved in string theory even classically. In many cases, the existence of extended objects -- string and branes -- and the degrees of freedom from them were essential to resolve the singularities. However, even though there is a common theme, details differ from case to case. Also, we have seen that some singularities should not be ``resolved" but must be ``excised" by the theory. So, it seems that there is no single mechanism which resolves all singularities. Probably, part of the problem is that there are many types of singularities. One cannot even define what is a singularity, and a singularity is defined indirectly through geodesic incompleteness. Moreover, even though various mechanisms are found to resolve singularities, it is not clear how to resolve the most important singularities -- the Schwarzschild and the big bang singularities.


Finally, it is important to note that singularities we covered are mainly BPS solutions. But it may be too optimistic to imagine that singularity resolutions for BPS cases also apply to non-BPS cases. Such a criticism was often made for the microscopic derivation of black hole entropy using D-branes, but a similar story applies here as well. As an example, we saw that the state which corresponds to the $M<0$ Schwarzschild should not exist at quantum level. In contrast, naked singularities in \sect{recent} are often various BPS configurations of branes, so they must exist at quantum level. Thus, those naked singularities had to be somehow regularized. Reference~\cite{Gubser:2000nd} also gives some argument that singularities must make sense if they can be obtained as the zero-temperature limit of black holes. 
This indicates that we cannot discuss a non-BPS naked singularity on the same footing as a BPS naked singularity. The upshot is that BPS solutions are very different from the real world, and there is no reason to believe that they have something in common.

\section*{Acknowledgments}

We would like to thank Gary Horowitz, Akio Hosoya, Veronika E. Hubeny, Akihiro Ishibashi, Gungwon Kang, David Kastor, Hideo Kodama, Donald Marolf, Kengo Maeda, Ken-ichi Nakao, Takashi Okamura, Masa-aki Sakagami, and Jiro Soda for useful discussions. I would also like to thank the participants and organizers of 
 workshops		
and of ``Working group on singularity in string and general relativity" for their questions and comments. This work was supported in part by the Grant-in-Aid for Scientific Research (13740167) from the Ministry of Education, Culture, Sports, Science and Technology, Japan.

\end{document}